\documentclass[a4paper,12pt]{article}
\usepackage{mathptmx}
\usepackage{epsfig}
\usepackage{float}
\usepackage{amssymb}
\usepackage{latexsym}
\usepackage{authblk}

\usepackage{amsfonts}
\newcommand{\beq}{\begin{equation}}
\newcommand{\eeq}{\end{equation}}
\newcommand{\beqa}{\begin{eqnarray}}
\newcommand{\eeqa}{\end{eqnarray}}

\usepackage{natbib}
\usepackage{graphicx}
\usepackage{multirow}
\begin{document}
%\begin{frontmatter}

%Gummi|065|=)
%\title{\textbf{The precipitation electron induced  ionization in the atmosphere:\\Two models of calculations: CRAC:EPII and AIMOS} }

%\maketitle

\title{Assessment of the Radiation Environment at Commercial Jet Flight Altitudes During GLE 72 on 10 September 2017 Using Neutron Monitor Data}
\author[1,2]{A.L. Mishev}
\author[1,2]{I.G. Usoskin}
\affil[1]{Space Climate Research Unit, University of Oulu, Finland.}
\affil[2]{Sodankyl\"a Geophysical Observatory, University of Oulu, Finland.}
\maketitle
\begin{abstract}
As a result of intense solar activity during the first ten days of September, a ground level enhancement occurred on September 10, 2017.  Here we  
computed the effective dose rates in the polar region at several altitudes during the event using the derived rigidity spectra of the energetic solar protons. The contribution of different populations of energetic particles viz. galactic cosmic rays and solar protons, to the exposure is explicitly considered and compared. We also assessed the exposure of a crew members/passengers to radiation at different locations and at several cruise flight altitudes and calculated the received doses for two typical intercontinental flights. The estimated received dose during a high-latitude, 40 kft, $\sim$ 10 h flight is $\sim$ 100 $\mu$Sv. 
\end{abstract}

\small Keywords:Solar eruptive events, Ground level enhancement, Neutron Monitor, Rigidity spectra of solar energetic protons,Exposure to radiation and received doses for crew members/passengers  
 \normalsize

\label{cor}{\small For contact: alexander.mishev@oulu.fi}

%\end{frontmatter}

\section{Introduction}

\begin{figure}
\centering \resizebox{\columnwidth}{!}{\includegraphics{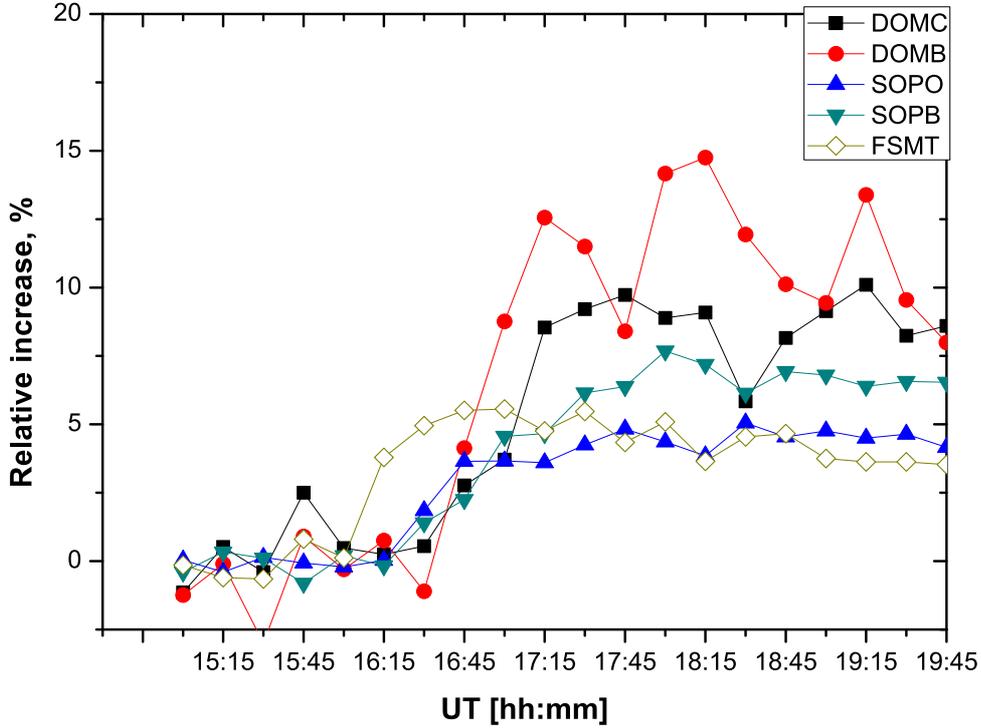}}
\caption{15-min averaged count rate variations of NMs with maximal increases during GLE 72 on September 10,  2017. The DOMC and SOPO correspond to standard NMs at Dome-C and South Pole stations, DOMB and SOPB correspond to the lead free NMs at at Dome-C and South Pole stations. Data are available at \texttt{http://gle.oulu.fi}.}
\label{Fig1}
\end{figure}

Intense solar activity took place during the first ten days of September 2017. This time period was among the most flare productive of the ongoing solar cycle 24.  The solar active region 12673  produced several X-class flares and coronal mass ejections (CMEs), leading to a moderate solar energetic particle (SEP) event, followed by a stronger, more energetic one, which was observed even at the ground level by several neutron monitors (NMs) (see the International GLE database \texttt{http://gle.oulu.fi}), i.e., the ground level enhancement (GLE) 72 event on September 10, 2017. The GLE 72 was related to an X8.2 solar flare, which peaked at 16:06 UT. It produced a gradual SEP event. At ground level, the event onset was observed at $\approx$ 16:15 UT (Fort Smith NM).  Records of NMs with maximal count rate increases during the event are shown in Fig.~\ref{Fig1}. The maximal count rate increases were observed by high-altitude standard and lead-free, i.e. without $Pb$ producer, monitors at Concordia station, 75.06 S, 123.20 E, 3233 m above sea level (a.s.l.), (DOMC/DOMB, 10--15 $\%$ above the pre-increase levels), South Pole  2820 m a.s.l. (SOPO/SOPB, 5--8 $\%$) and at the sea level Forth Smith - FSMT ($\approx$ 6 $\%$). The lead free NMs (DOMB and SOPB) are more sensitive compared to standard NMs. In addition, high-altitude NMs are more sensitive than sea level NMs.

Strong SEP events can significantly change the radiation environment in the vicinity of Earth and in the Earth's polar atmosphere, where the magnetospheric shielding is marginal \citep[e.g.][and references therein]{Spurny02, Vainio2009}. While cosmic rays (CRs) of galactic origin permanently govern the radiation environment in the global atmosphere, particles of  solar origin, specifically during strong SEP and GLE events can considerably enhance the flux of secondary CR particles in the atmosphere. Primary CR particles penetrate into the atmosphere and induce a complicated nuclear-electromagnetic-muon cascade, producing large amount of various types of secondary particles, viz. neutrons, protons,
 $\gamma$, $e^{-}$, $e^{+}$, $\mu^{-}$, $\mu^{+}$, $\pi^{-}$, $\pi^{+}$, distributed in a wide energy range, which eventually deposit their energy and ionize the ambient air \citep{Bazilevskaya08, Asorey2018461}. Hence, CR particles determine the complex radiation field at flight altitudes \citep{Spurny96, Shea2000}. 

Assessment of the radiation exposure, henceforth exposure, at typical flight altitudes is an important topic in the field of space weather \citep[e.g.][and references therein]{Baker1998, Latocha2009286, Lilensten2009, Mertens2013, Mertens2016}. Individual accumulated doses of the cockpit and cabin crew are monitored and crew members are regarded as occupational workers \citep{ICRP2007, Euratom13}. The contribution of galactic cosmic rays (GCRs) to the exposure can be assessed by computations and/or using corresponding data sets for solar modulation and reference data \citep[e.g.][and references therein]{Menzel2010, Meier2018}, considering explicitly the altitude, geographic position, solar activity, geomagnetic conditions \citep{Spurny02, Shea2000, Tobiska2018}. On the other hand, the assessment of exposure during GLEs can be rather complicated, because of their sporadic occurrence and a large variability of their spectra, angular distributions, durations and dynamics \citep{Gopalswamy201223, Moraal201285}. For a precise computation of the exposure during a GLE event, it is necessary to possess appropriate information about the energy and angular distribution of the incoming high-energy particles \citep{Kuwabara2006}. Such computations are performed on a case-by-case basis for individual events \citep[e.g.][]{Sato2018924}. 

Here, we computed the effective dose rates during GLE 72 at several cruise flight altitudes. We employed a recently developed model and procedure, the details are given in \citet{Mishev2015} and \citet{Mishev2017swsc}. We calculated the exposure over the globe and the received doses of crew members/passengers for typical intercontinental flights. 

\section{Reconstruction of proton spectra for GLE 72 using NM data}
Using a model briefly described below and actual records from the global NM network, we derived the rigidity spectra and angular distributions of solar protons for GLE 72, see details in \citet{Mishev2018sol72}. Estimates of GLE characteristics viz. rigidity/energy spectra and angular distributions can be performed using the NM data and a corresponding model of the global NM network response \citep[e.g.][]{Shea82, Cramp97}. In this study we employed a method described in great detail elsewhere \citep{Mishev14c, Mishev16SF}. Modelling of the global NM response was performed using a recently computed NM yield function \citep{Mishev13b, Gil15, Mangeard201611}, which results in an improved convergence and precision of the optimization \citep{Mishev2017swsc}.  

Here, we assume the rigidity spectrum of the GLE particles to be a modified power law similar to \citet{Vas08}:

\begin{equation}
    J_{||}(P)= J_{0}P^{-(\gamma+\delta\gamma(P-1))}
        \label{simp_eqn1}
   \end{equation}

\noindent where $J_{||}(P)$ is the differential flux of solar particles with a given rigidity $P$ in [GV] arriving from the Sun along the axis of symmetry, whose direction is defined by the geographic coordinates $\Psi$ (latitude) and $\Lambda$ (longitude), $\gamma$ is the power-law spectral exponent and  $\delta\gamma$ is the corresponding rate of steepening of the spectrum. The pitch-angle distribution (PAD) is assumed to be a superposition of two oppositely directed (Sun and anti-Sun) Gaussians: 

\begin{equation}
        G(\alpha) \sim \exp(-\alpha^{2}/\sigma_{1}^{2}) + B*\exp(-(\alpha-\pi)^{2}/\sigma_{2}^{2}) 
        \label{simp_eqn2}
   \end{equation}

\noindent where $\alpha$ is the pitch angle, i.e., the angle between the charged particle's velocity vector and the local magnetic field direction, $\sigma_{1}$ and $\sigma_{2}$ are parameters corresponding to the width of the PAD, and B corresponds to the contribution of the particle flux arriving from the anti-Sun direction. 

\begin{figure}
\centering \resizebox{\columnwidth}{!}{\includegraphics{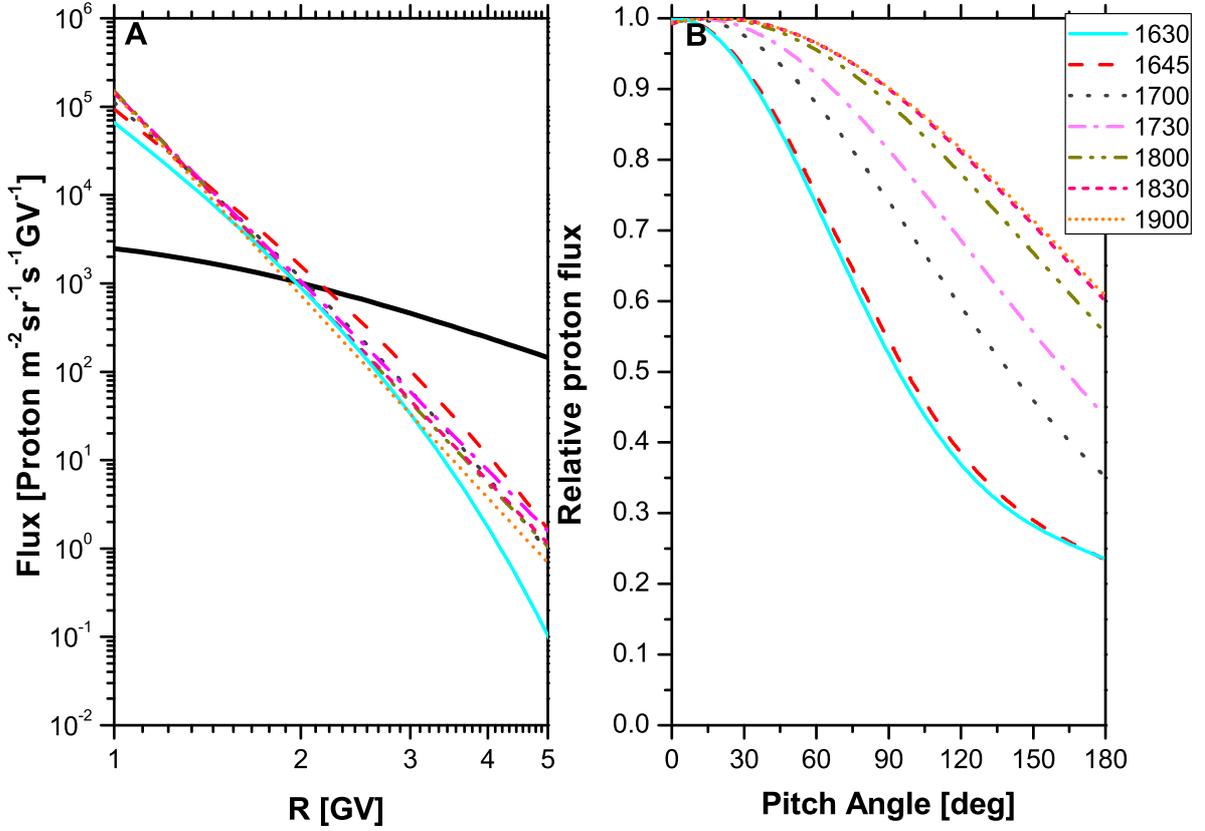}}
\caption{GLE particles rigidity  spectra and PAD during GLE 72 on September 10, 2017, details are given in Table 1. Time (UT) corresponds to the start of the five minute interval over which the data are integrated. The black solid line of the left panel denotes the GCR particle flux computed on period corresponding to GLE 72 occurrence.}
\label{Fig2}
\end{figure}      

The rigidity spectrum and PAD are derived by minimizing the functional form $\mathcal{F}$ which is the sum of squared differences between the model $\frac{\Delta N_{i}}{N_{i}}_{mod.}$ and measured $\frac{\Delta N_{i}}{N_{i}}_{exp.}$ relative increases of NMs:

\begin{equation}
\mathcal{F}=\sum_{i=1}^{m} \left[\left(\frac{\Delta N_{i}}{N_{i}}\right)_{mod.}-\left(\frac{\Delta N_{i}}{N_{i}}\right)_{exp.}\right]^{2}
\label{simp_eqn3}
   \end{equation}

\noindent over $m$ NM stations, where $\Delta N_{i}$ and $N_{i}$ are the the count rate increase due to solar protons and the pre-event background counts due to GCRs of the $i$-th NM, respectively. Herein, the minimization of $\mathcal{F}$ is performed using a variable regularization similar to that proposed by \citet{Tikhonov1995} employing the Levenberg-Marquardt method \citep{Lev44, Mar63}. The goodness of the fit is based on residual $\mathcal{D}$ (Equation 4) \citep[e.g.][]{Himmelblau72, Den83}.

\begin{equation}
\mathcal{D}=\frac{\sqrt{\sum_{i=1}^{m} \left[\left(\frac{\Delta N_{i}}{N_{i}}\right)_{mod.}-\left(\frac{\Delta N_{i}}{N_{i}}\right)_{meas.}\right]^{2}}}{\sum_{i=1}^{m} (\frac{\Delta N_{i}}{N_{i}})_{meas.}}
\label{simp_eqn4}
   \end{equation} 
 
During the analysis, the background due to GCRs was averaged over two hours before the event's onset, and the Forbush decrease started, on September 7, 2017,  was explicitly considered in our analysis. Here we present the derived SEP characteristics, expanding the time interval reported in \citet{Mishev2018sol72}. The derived rigidity spectra of GLE particles were found to be relatively hard  during the event onset (see Fig.2a) for a weak event and a softening of the spectra throughout the event was derived \citep[e.g.][]{Mishev2017swsc, Mishev2018sol72}. The derived spectral index after the event onset is in very good agreement with other estimates \citep[e.g.][]{Kataoka2018}. After 17:15 UT the energy distribution of the GLE particles was described by a pure power-law rigidity spectrum. In addition, it was recently shown that this event was softer at high energies than average GLEs, but revealed hard spectrum at low energies \citep[e.g.][]{Cohen2018}. The angular distribution of the high energy solar particles broadened out throughout the event and was wide, except for the event onset (see Fig.2b). We assumed an isotropic SEP flux for conservative assessment of the exposure similarly to \citet{Copeland20081008}. The derived spectra and angular distributions will be integrated into the GLE database \citep{Tuohino2018398}.

\section{Assessment of effective dose rate at aviation altitudes during GLE 72}
For the calculation of the effective dose rates during GLE 72 we employed a recently developed numerical model, which is based on pre-computed effective dose yield functions from high-statistics Monte Carlo simulations. These yield functions are the response of ambient air at a given altitude $h$ above sea level as the effective dose to a mono-energetic unit flux of primary CR particle entering the Earth's atmosphere. 

The effective dose rate at a given atmospheric altitude $h$ a.s.l. induced by primary CR particles is given by the expression:	
	
\begin{equation}\label{eq:1}
E(h, T, \theta, \varphi) = \sum_{i}\int_{T_{cut,i}(P_{cut})}^{\infty}\int_{\Omega}J_i(T)Y_i(T, h)d\Omega(\theta, \varphi) dT,
\label{simp_eqn5}
\end{equation}

\noindent where $P_{cut}$ is the local geomagnetic cut-off rigidity, $\Omega$ is a solid angle determined by the angles of incidence of the arriving particle $\theta$ (zenith) and $\varphi$ (azimuth),  $J_{i}(T)$ is the differential energy spectrum of the primary CR at the top of the atmosphere for nuclei of type $i$ (proton or $\alpha-$particle) and $Y_{i}$ is the corresponding yield function. The integration is over the kinetic energy above $T_{cut,i}(P_{cut})$, which is defined by  $P_{cut}$ for a nuclei of type $i$. The full description of the model with the corresponding look-up tables of the yield functions at several altitudes a.s.l. and comparison with reference data is given elsewhere \citep{Mishev2015}. 

Here we computed the effective dose rate during GLE 72 using newly derived SEP spectra and angular distributions on the basis of NM data (details are given in Section 2) and Eq. 5. The exposure during GLE events is defined as a superposition of the GCRs and SEPs contributions. The radiation background due to GCR was computed by applying the force field model of galactic cosmic ray spectrum \citep{Gle68, Bur00, Uso05} with the corresponding parametrization of local interstellar spectrum \citep[e.g.][]{Uso05, Usos06}, where the modulation potential is considered similar to \citet{Usoskin11a}. For the computation of the exposure we do not consider the depression of GCRs due to the Forbush decrease, started on September 7, 2017. This results in a conservative approach for the contribution of GCRs to the exposure with eventual overestimation of the background exposure. Accordingly, the characteristics of energetic solar protons used in Eq. 5 were taken from Fig.1. The flux of incoming GLE particles was assumed to be isotropic, which is consistent with the derived angular distribution and allows one to assess conservatively the exposure \citep[e.g.][]{Copeland20081008}. 

\begin{figure}
\centering \resizebox{\columnwidth}{!}{\includegraphics{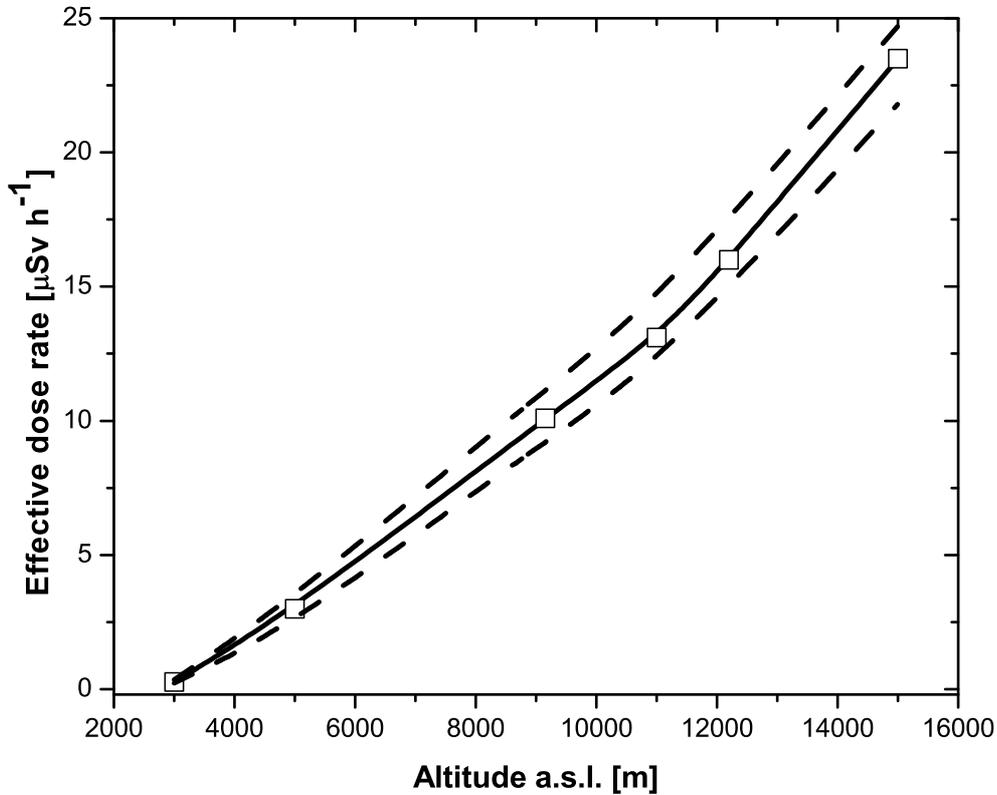}}
\caption{Computed maximal effective dose rate as a function of altitude a.s.l. during the main phase of GLE 72 on September 10, 2017. The dashed lines encompass the 95 $\%$ confidence interval}. 
\label{Fig4}
\end{figure}  

In this way we computed the effective dose rate during GLE 72 at several typical for cruise flight altitudes, namely 30 kft (9 100m), 35 kft (10 670m), 40 kft (12 200m) and 50 kft (15 200m) a.s.l.. The effective dose rate was estimated also at high-mountain altitude of about 3000 and 5000 m a.s.l. using the yield functions by \citet{Mishev201615}. These computations were performed for a high-latitude region with a low cut-off rigidity $P_{cut}$ $<$ 1 GV, where the expected exposure is maximal. Results during period with maximum exposure are presented in Fig.~\ref{Fig4}. The time evolution of the exposure throughout the event is computed at several altitudes. 

One can see that the contribution of SEPs to the total exposure is comparable to the contribution due to GCRs, except for low altitudes. At the ground level, the contribution of SEPs to the total exposure is small, because of their considerably softer spectrum, compared to GCRs. The peak exposure is in the range of 20--24 $\mu$Sv.h$^{-1}$ at altitude of 50 kft a.s.l., 11--13 $\mu$Sv.h$^{-1}$ at altitude of 35 kft a.s.l. and about 10 $\mu$Sv.h$^{-1}$ at altitude of 30 kft a.s.l., during the main phase of the event, i.e., between 17:00 and 18:30 UT. During the late phase of the event (after 21:00 UT), the exposure decreases to roughly 20 $\mu$Sv.h$^{-1}$, 12 $\mu$Sv.h$^{-1}$ and about 10 $\mu$Sv.h$^{-1}$ at altitudes of 50, 35 and 30 kft a.s.l., respectively. The contribution of solar protons to the exposure considerably decreases during the late phase of the event.

\begin{figure}
\centering \resizebox{\columnwidth}{!}{\includegraphics{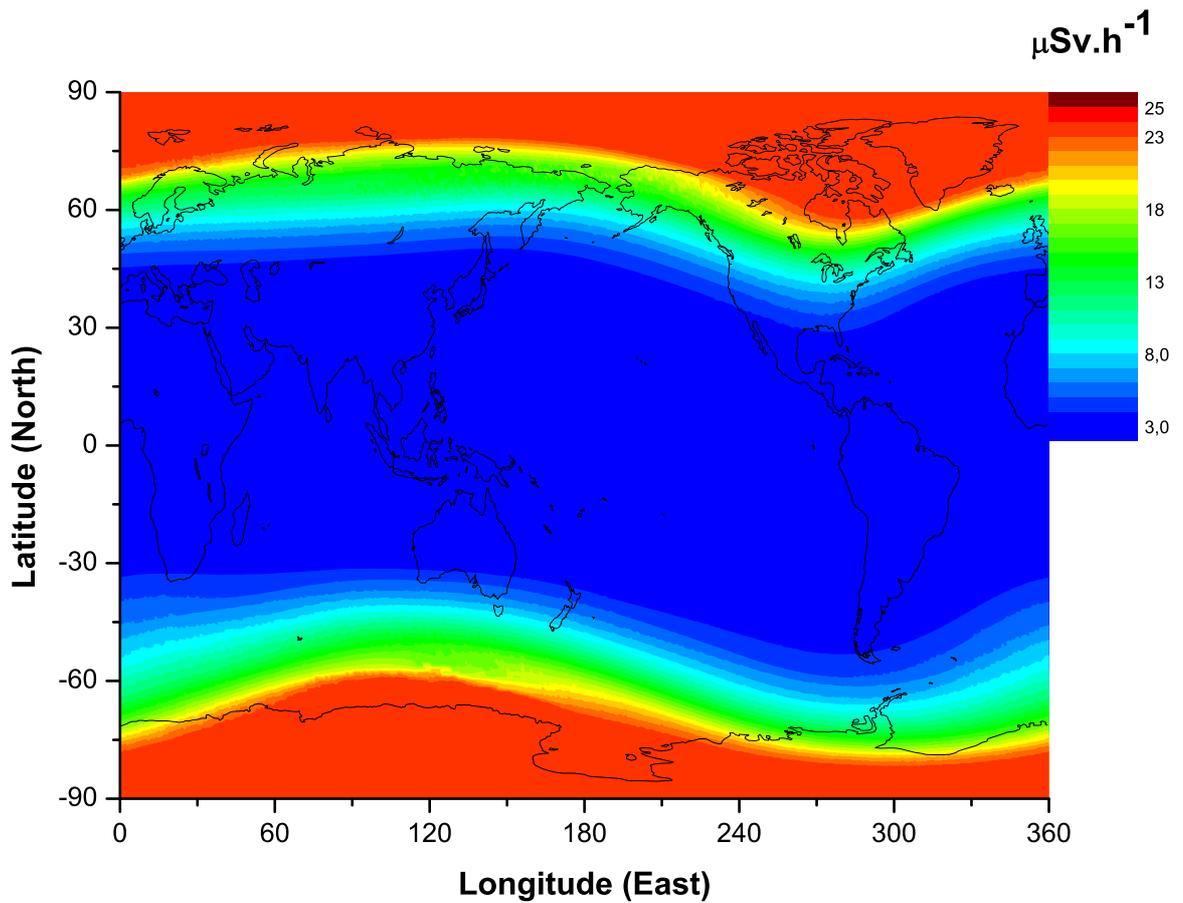}}
\caption{Distribution of the effective dose rate as a function of the geographic coordinates at altitude of 50 kft due to high energy GLE and GCR particles during the main phase of GLE 72 on September 10, 2017.}
\label{Fig5}
\end{figure}

\begin{figure}
\centering \resizebox{\columnwidth}{!}{\includegraphics{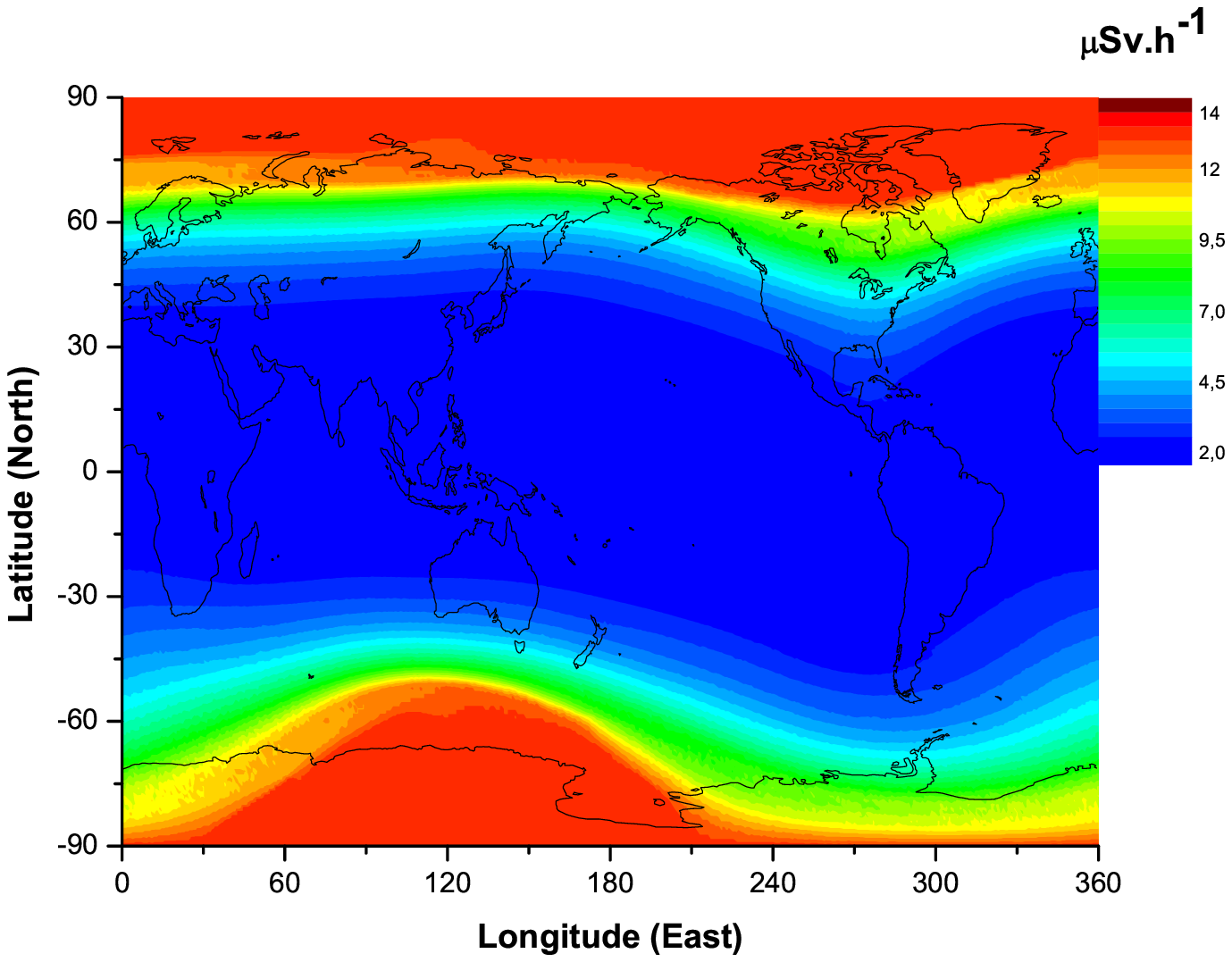}}
\caption{Distribution of the effective dose rate as a function of the geographic coordinates at altitude of 35 kft due to high energy GLE and GCR particles during the main phase of GLE 72 on September 10, 2017.}
\label{Fig6}
\end{figure}

\begin{figure}
\centering \resizebox{\columnwidth}{!}{\includegraphics{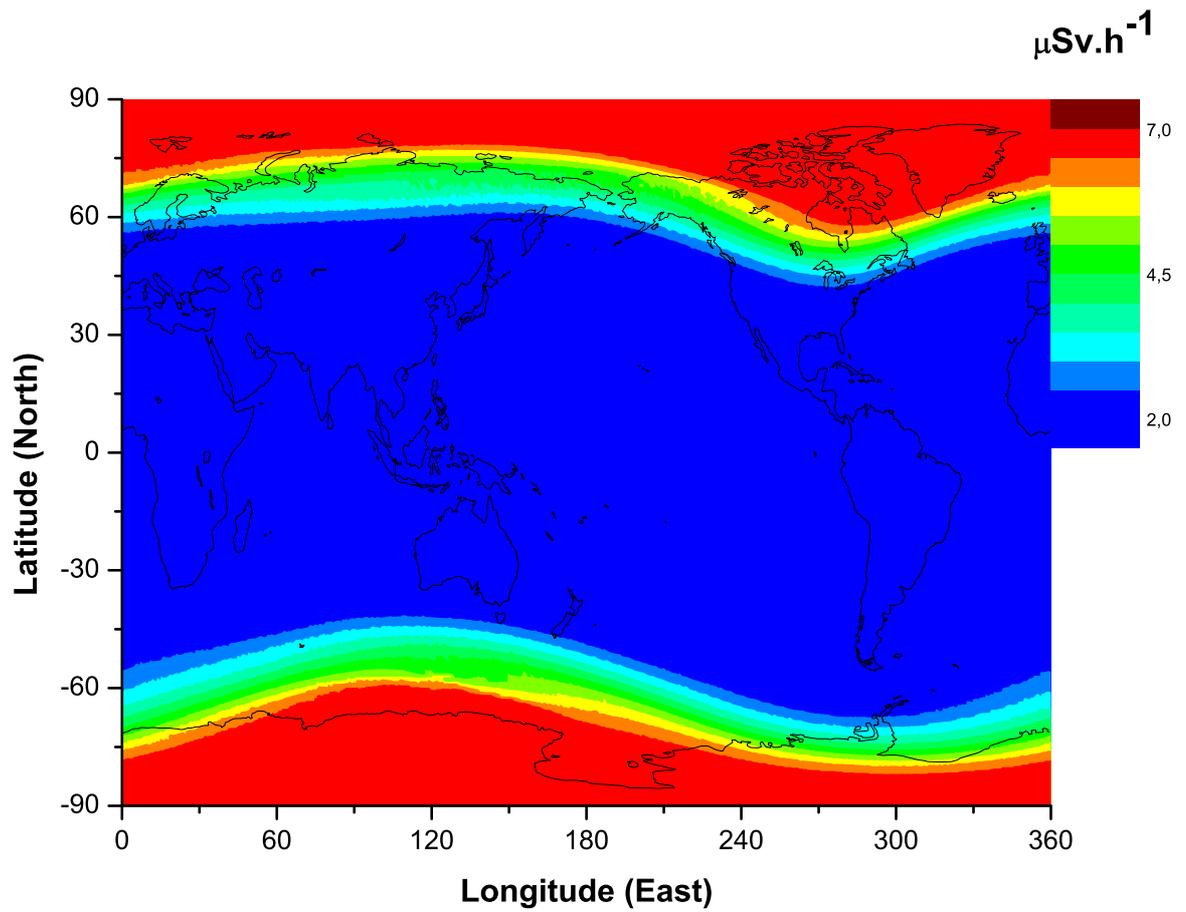}}
\caption{Distribution of the effective dose rate as a function of the geographic coordinates at altitude of 30 kft due to high energy GLE and GCR particles during the main phase of GLE 72 on September 10, 2017.}
\label{Fig7}
\end{figure}

The distribution of the exposure over the globe is determined by the cut-off rigidity, which is computed here using a combination of Tsyganenko 1989 (external) \citep{Tsyganenko89} and IGRF (internal) \citep{Lan87} geomagnetic models. This combination allows one to compute straightforwardly the cut-off rigidity with a reasonable precision \citep{Kudela04, Kud08, Nevalainen13}. An example of the distribution of the exposure as a function of the geographic coordinates for altitude of 50 kft a.s.l. during the main phase of GLE 72 is given in Fig.~\ref{Fig5}. The distribution of the effective dose rate reveals a maximum at polar and sub-polar regions and rapidly decreases at regions with higher cut-off rigidity. Similar computations were performed for lower cruise flight altitudes, the results are presented  in Fig.~\ref{Fig6} (35 kft a.s.l.) and Fig.~\ref{Fig7} (30 kft a.s.l.). Computations for the late phase of the event depict similar distributions of the exposure, but with lower values.
Those results are valid for the polar regions, while at low latitudes there is no notable change of the expected exposure, which is due to GCRs. Moreover, even a slight increase of the exposure at low latitudes is expected, because of the recovery of the Forbush decrease, but not considered here.

The exposure decreases significantly as a function of increasing cut-off rigidity. Below 30 kft, as well as at regions with $P_{cut}$ $\geq$ 2 GV,  the contribution of SEPs becomes small even negligible, because their spectrum is considerably softer than the GCR spectrum. 

The computed distributions of effective dose rates allow one to estimate the exposure of a crew members/passengers on board of a transcontinental flight during the GLE 72. Here we consider nearly a worst-case scenario, i.e., a polar route, departure time close to the event onset, high constant cruise altitude of 40 kft and a conservative approach for the exposure by assuming an isotropic SEP flux, without considering the effect of the Forbush decrease. Therefore, we present a very conservative assessment of the received effective dose by crew members/passengers during the GLE 72.
 
As an example, a crew members/passengers, would receive about 90 $\mu$Sv on a flight from Helsinki (HEL), Finland  to Osaka (KIX), Japan (departure time 17:10 UT, 9h 30m duration, altitude 40 kft), 110 $\mu$Sv from Helsinki to New York--JFK (departure time 15:20 UT, 8h 40m duration, altitude 40 kft), respectively. Here, We do not consider change of the flight altitude during the ascending and the landing phase in order to conservatively assess the exposure. In both cases, the flight routes are along the great circle. Despite the shorter HEL--JFK flight, one would receive larger exposure, mostly because of the polar route. In addition, the  HEL--JFK flight is during the main phase of the event, while HEL--KIX flight is during the main and late phase of the event, because of the later departure, according the actual flight information.

These results related to radiation environment during GLE 72 are compared with other similar estimates \citep[e.g.][]{Copeland2018, Kataoka2018, Matthia2018a}. A good agreement, in the order of 10--14 $\%$, at altitude of 50 kft with the exposure reported by \citet{Copeland2018} is achieved. At lower levels the difference increases to 40--55 $\%$ at altitude of 40 kft and to 75 $\%$ at altitude of 35 kft, respectively. In all cases our model reveals greater exposure. The differences are consistent with recent reports  \citep[e.g.][]{But13, Butikofer2015}. They are most likely due to the slightly different SEP spectra derived using NM data (our analysis), compared to GOES data analysis \citep[e.g.][]{Copeland2018}. 

\section{Summary and Discussion}
In this study we presented reconstruction of rigidity spectrum and PAD of solar energetic protons during the GLE 72 using data from the global neutron monitor network. Using the reconstructed spectrum we assessed the exposure for crew members/passengers at several typical cruise flight altitudes in a polar region, assuming a conservative isotropic approach of the GLE particles angular distribution. We also conservatively calculated the received doses for two typical intercontinental flights: HEL-- KIX (departure time 17:10 UT, 9h 30m duration, altitude 40 kft) and HEL--JFK (departure time 15:20 UT, 8h 40m duration, altitude 40 kft). We conclude that during a weak GLE event such as GLE 72 on September 10, 2017, the upper limit of the radiation exposure over a single flight is about 100 $\mu$Sv, with contribution of GCRs of about 60--65 $\mu$Sv, does not represent an important space weather issue. Usually, the pilots receive annually more than the annual general public limit of 1 mSv \citep[e.g.][]{Euratom13}, with the majority receiving around 3 mSv \citep[e.g.][]{Bennett2013b}. However, the exposure during GLEs should be monitored. The presented results can be compared with other similar estimates. 

The exposure at cruise flight altitudes during strong SEP events can be significantly enhanced compared to quiet periods. It is a superposition of contributions of GCRs and SEPs. As a result, during strong SEP events and GLEs, crew members/passengers may receive doses well above the background level due to GCRs  \citep[e.g.][]{Matthia2009b, Tuohino2018398}. While the background exposure due to GCRs can be assessed by computations and/or on the basis of appropriate measurements, the estimation of the exposure due to high energy SEPs is rather complicated and it is performed retrospectively. Occurring sporadically, GLEs differ from each other in spectra and duration, and are therefore usually studied case by case. Deep and systematic study of the exposure during GLEs provides a good basis for further assessment of space weather effects related to accumulated doses at aviation flight altitudes and allows one to compare and adjust possible uncertainties in the existing methods and models in this field. 

%%%%%%%%%%%%%%%%%%%%%%%%%%%%%%%%%%%%%%%%%%%%%%%%%%%%%%%%%%%%%%%%%%%%%%%%%%%
\section*{Acknowledgements}
This work was supported by the Academy of Finland (project No. 272157, Center of Excellence ReSoLVE and project No. 267186). French-Italian Concordia Station (IPEV program n903 and PNRA Project LTCPAA PNRA14 00091) is acknowledged for support of DOMC/DOMB stations as well as the projects CRIPA and CRIPA-X No. 304435 and Finnish Antarctic Research Program (FINNARP). We acknowledge neutron monitor database (NMDB) and all the colleagues and PIs from the neutron monitor stations, who kindly provided the data used in this analysis, namely: Alma Ata, Apatity, Athens, Baksan, Dome C, Dourbes, Forth Smith, Inuvik, Irkutsk, Jang Bogo, Jungfraujoch, Kerguelen, Lomnicky \v{S}tit,  Magadan, Mawson, Mexico city, Moscow, Nain, Newark, Oulu, Peawanuck, Potchefstroom, Rome, South Pole, Terre Adelie, Thule, Tixie. The NM data are available on-line at International GLE database \texttt{http://gle.oulu.fi}.

%\acknowledgment US spelling: \verb+\acknowledgment+
%\acknowledgement British  spelling: \verb+\acknowledgement+

%%%%%%%%%%%%%%%%%%%%%%%%%%%%%%%%%%%%%%%%%%%%%%%%%%%%%%%%%%%%%%%%%%%%%%%%%%%

%\end{article} 

\end{document}